\documentclass{webofc}
\usepackage[varg]{txfonts}   

\usepackage{lineno}
\newcommand{\hlt}{^{3}_{\Lambda}\mathrm{H}}
\newcommand{\hlf}{^{4}_{\Lambda}\mathrm{H}}
\newcommand{\hel}{^{4}_{\Lambda}\mathrm{He}}
\newcommand{\snn}{\sqrt{s_{\mathrm{_{NN}}}}}

\begin{document}
%
\title{Measurements on the production and properties of light hypernuclei at STAR}
%
%

\author{\firstname{Yuanjing} \lastname{Ji}\inst{1}\thanks{\email{yuanjingji@lbl.gov}}, for the STAR Collaboration
}

\institute{Lawrence Berkeley National Laboratory}

\abstract{%
  The hyperon-nucleon ($Y$-$N$) interaction, an important ingredient for the nuclear equation-of-state (EoS), remains poorly constrained. 
  Precise measurements of hypernucei intrinsic properties and production yields in heavy-ion collisions are crucial for the understanding of their production mechanisms and the strength of the $Y$-$N$ interaction. 
  Thanks to the high statistics data taken from the STAR BES II program,  a series of hypernuclei measurements are carried out at low energies. 
  In these proceedings, we present the kinematic and centrality dependence of light hypernuclei production yields and strangeness population factor ($S_3$, $S_4$) in Au+Au collisions at $\sqrt{s_{\rm NN}}=$ 3 GeV. We also report the energy dependence of $^{3}_{\Lambda}{\rm H}$ yields and $S_3$ at mid-rapidity from 3 to 27 GeV Au+Au collisions. Precise measurements of $^{4}_{\Lambda}{\rm He}$ lifetime and $^{3}_{\Lambda}{\rm H}$ branching ratio are also reported. These results are compared with model calculations and physics implications are discussed.
}
\maketitle
\section{Introduction}
\label{sec:intro}
Hypernuclei are bound systems of nucleons and hyperons. They introduce an additional degree of freedom in baryon interactions from hyperons. Thus, hypernuclei are regarded as natural laboratory to investigate hyperon-nucleon ($Y$-$N$) interactions. $Y$-$N$ interaction is the important ingredient for understanding the Equation of State (EoS) of neutron stars and the hadronic phase of heavy-ion collisions. Thermal models \cite{Andronic:2010qu} predict that hypernuclei are abundantly produced in heavy-ion collisions at high baryon density regions. The second phase of the Beam Energy Scan at RHIC, BES-II program, collects high statistical data in Au+Au collisions with center of mass energies ranging from 3-27 GeV. 
It maps the QCD phase diagram from 200 MeV baryon chemical potential ($\mu_B$) to around 750 MeV. Thus, BES II program provides us an excellent opportunity to investigate hypernuclei physics in heavy-ion collisions. 

\section{Analysis details}
\label{sec:exp}
For low-energy collisions (3-17 GeV), the collider can be run under the Fixed-target (FXT) mode to significantly enhance the luminosity. Around $2.6\times10^{8}$ good events are collected for 3 GeV Au+Au collisions in 2018. 
The gold target is located at the west side of the STAR detector, while the gold beam comes from the west to east direction.
Tracks are reconstructed by Time Projection Chamber (TPC), which covers the pseudo-rapidity range of $-1.5<\eta<0$ with respect to the target position. Charged particles are identified utilizing the particle energy loss information provided by TPC. 
In these proceedings, $\hlt$ is reconstructed via both $\hlt\rightarrow dp\pi^{-}$ and $\hlt\rightarrow {\rm ^{3}He}\pi^{-}$ channels, and $\hlf$ and $\hel$ are reconstructed via $\hlf\rightarrow {\rm ^{4}He}\pi^{-}$ and $\hel\rightarrow {\rm ^{3}He}p\pi^{-}$. Hypernuclei candidates are reconstructed utilizing KF Particle package \cite{Zyzak:2016exl} to enhance significance. 

\section{Results and discussions}
\subsection{Production of hypernuclei in heavy-ion collisions}
\label{sec2}
\begin{figure}[h]
\centering
\begin{minipage}{0.66\textwidth}
\centering
\includegraphics[height=4.1cm]{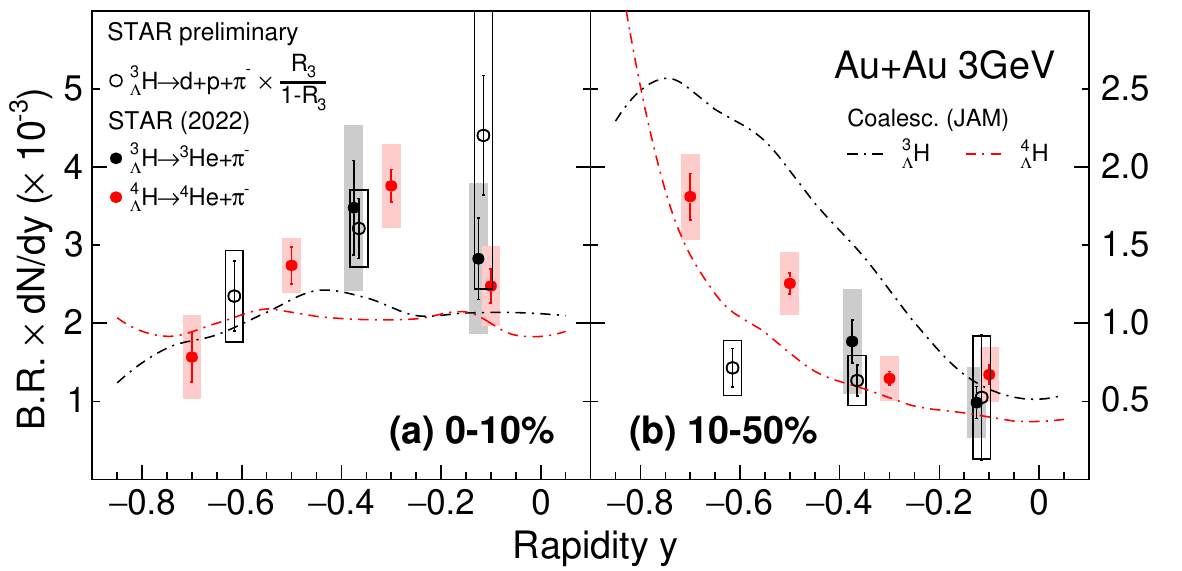}
\end{minipage}
\begin{minipage}{0.33\textwidth}
\centering
\includegraphics[height=4.1cm]{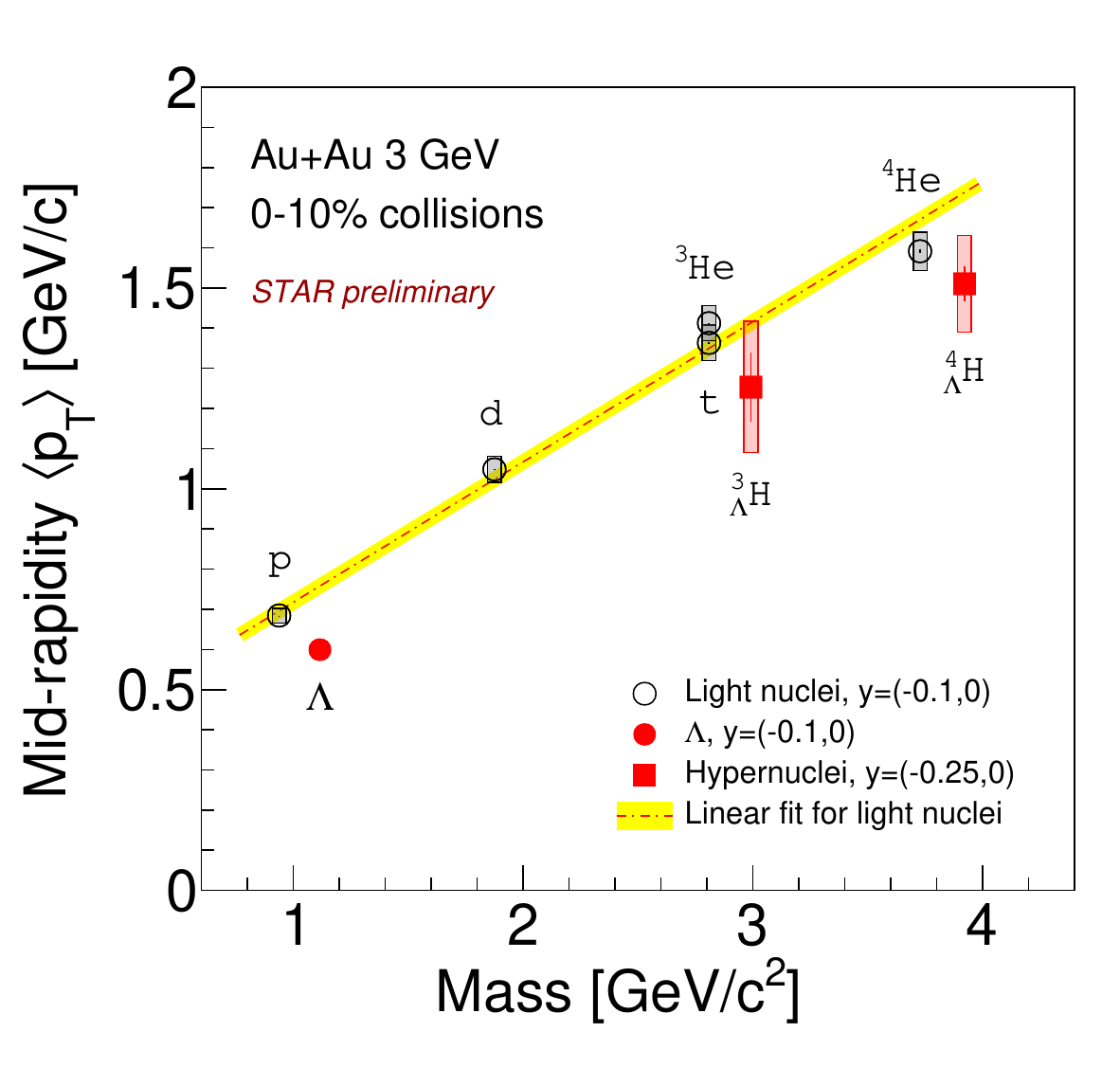}
\end{minipage}
\caption{The left plot shows the rapidity dependence of $\hlt$ (black solid and open circles), $\hlf$ (red solid circles) $B.R.\times dN/dy$ as a function of rapidity in $0$-$10\%$ and $10$-$50\%$ centralities. $\hlt$ yields via $\hlt\rightarrow dp\pi^{-}$ channel are scaled by a factor of $R_3/(1-R_3)$. Dashed lines are JAM calculations coupled with instant coalescence \cite{Liu:2019nii}. The right plot shows the average transverse momentum $\langle p_T \rangle$ of different particle species as a function of particle mass at mid-rapidity in 3 GeV Au+Au collisions.} 
\label{fig:h34ldndy}
\end{figure}

\begin{figure}
\centering
\sidecaption
\includegraphics[height=4.1cm]{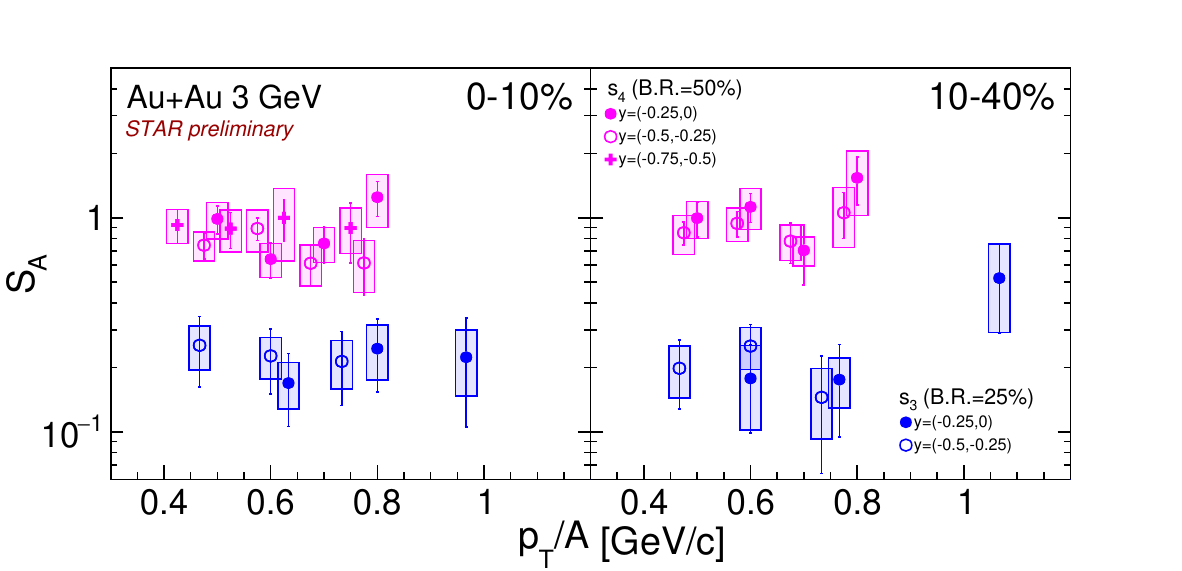}
\caption{$S_3$ (blue markers) and $S_4$ (magenta markers) are shown as a function of $p_T/A$, where $A$ is the mass number. $S_3$ and $S_4$ are presented in 3 rapidity bins at $0$-$10\%$ and $10$-$40\%$ centralities.}
\label{fig:s34_diff}
\end{figure}

Figure \ref{fig:h34ldndy} (left) shows the rapidity dependence of $\hlt$ and $\hlf$ production yields in Au+Au collisions at 3 GeV \cite{STAR:2021orx}. Both $\hlt$ and $\hlf$ show different trends in $0$-$10\%$ and $10$-$50\%$ centralities. Dashed lines are calculations from the transport model (JAM) with instant coalescence of all hadrons as an afterburner \cite{Liu:2019nii}. This simple coalescence model can qualitatively describe the data with tuned coalescence parameters. Figure \ref{fig:h34ldndy} (right) shows the light nuclei and hypernuclei average transverse momentum $\langle p_T\rangle$ as a function of particle mass. A linear trend is observed, suggesting the dominance of collective radial motion. Similar phenomena are also observed in light nuclei and hypernuclei directed flow measurements \cite{Hu:2022cpp}. Those results are qualitatively consistent with that hypernuclei are produced from the coalescence of hyperons and nucleons. 
To investigate the role of $Y$-$N$ interaction in heavy-ion collisions, 
we calculate the strangeness population factor \cite{Zhang:2009ba} that takes $\Lambda$ baryon over proton yield ratio as a reference to hypernuclei to light nuclei yield ratio:
\begin{equation}
S_A=\frac{^{A}_{\Lambda}\mathrm{H}(A\times p_T)}{^{A}\mathrm{He}(p_T)\times\frac{\Lambda}{p}(p_T)}=\frac{B_A(^{A}_{\Lambda}\mathrm{H})(p_T)}{B_A(^{A}\mathrm{He})(p_T)}
\label{eq:sa}
\end{equation}
As one can see in Eq. \ref{eq:sa}, $S_A$ has a direct connection with the ratio of coalescence parameters $B_{A}$. Figure \ref{fig:s34_diff} shows $S_A(A=3,4)$ as a function $p_T/A$ in $0$-$10\%$ and $10$-$40\%$ centralities, where $A$ is the mass number. Within the current uncertainties, no obvious $p_T$, rapidity and centrality dependence of $S_A$ are observed at 3 GeV. The results imply that $B_A$ of light nuclei and hypernuclei might follow similar tendency versus $p_T$, rapidity and centralities. Suppressed $\hlt/^{3}\mathrm{He}$ yield ratios are observed with respect to $\Lambda/p$ yield ratio in Au+Au collisions at 3 GeV with $S_3<1$. 
$\hlf/^{4}\mathrm{He}$ yield ratio is comparable to $\Lambda/p$ yield ratio, which can be explained by the feed-down contributions from the exited state $^{4}_{\Lambda}\mathrm{H^{*}}$($J^{+}=1$).

\begin{figure}
\begin{minipage}{0.495\textwidth}
\centering
\includegraphics[height=5.3cm]{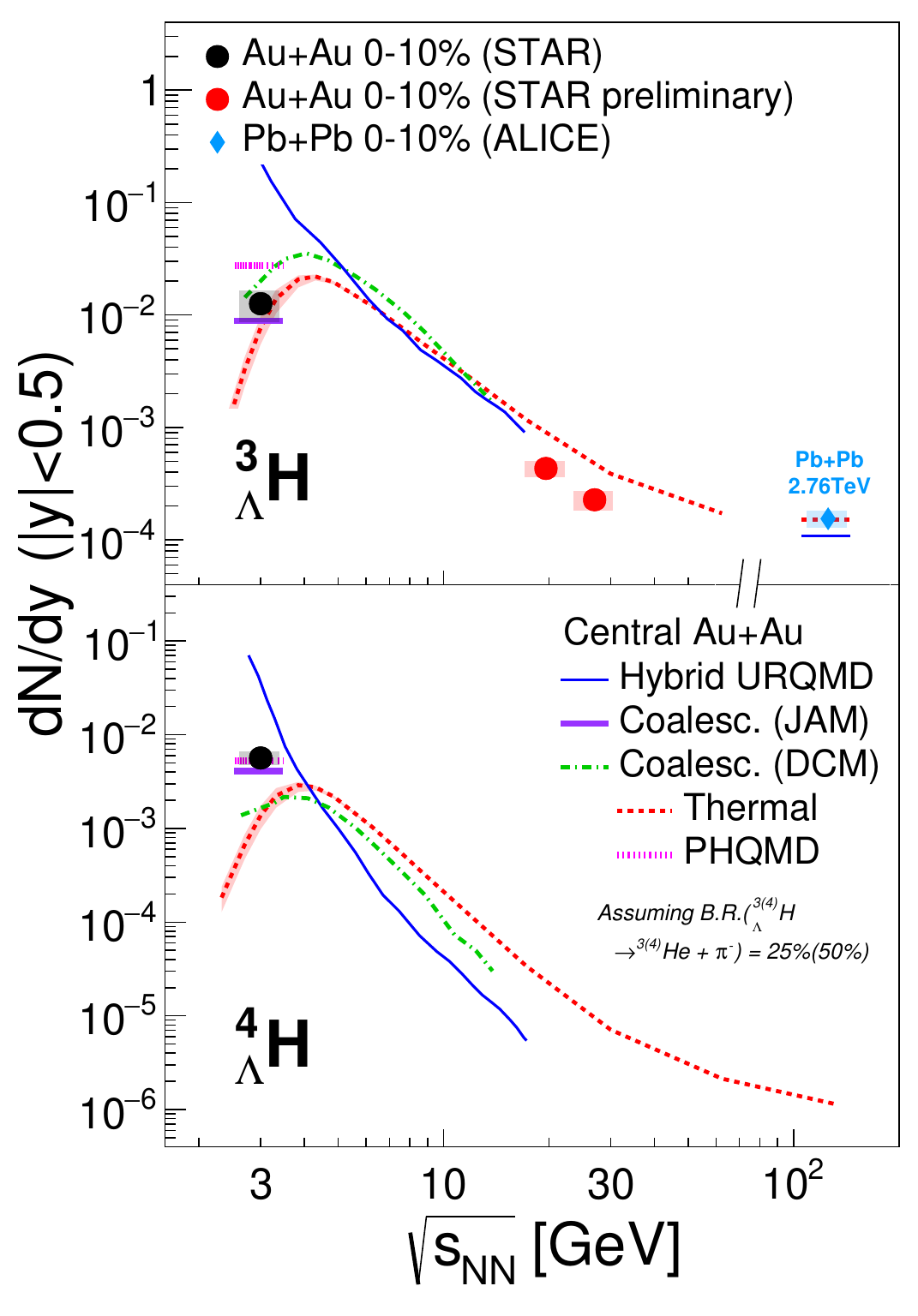}
\caption{$dN/dy$ of $\hlt$ (top panel) and $\hlf$ (bottom panel) at mid-rapidity as a function of the center-of-mass energy. Our results are compared with the following model calculations: Hybrid UrQMD \cite{Steinheimer:2012tb}, JAM \cite{Liu:2019nii}, DCM \cite{Steinheimer:2012tb}, Thermal \cite{Andronic:2010qu}, PHQMD\cite{Glassel:2021rod}. }
\label{fig:yields_energy}
\end{minipage}
\hspace{0.15cm}
\begin{minipage}{0.495\textwidth}
\vspace{0.6cm}
\includegraphics[height=4.2cm]{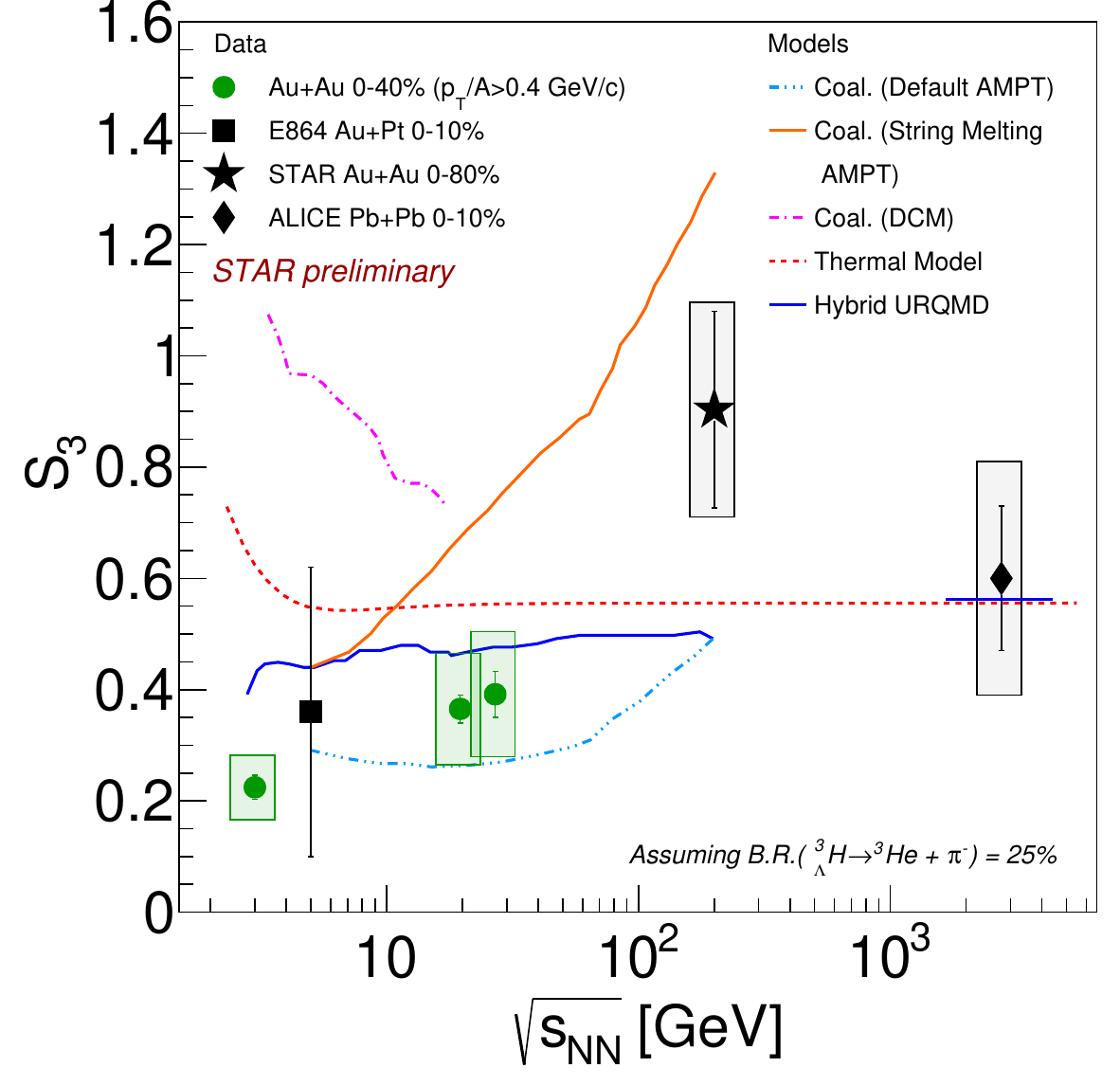}
\vspace{0.75cm}
\caption{The energy dependence of $S_3$ is shown. Green circles are the STAR new measurements. Model calculations shown in the plot are from: AMPT \cite{Zhang:2009ba}, DCM \cite{Steinheimer:2012tb}, Thermal \cite{Andronic:2010qu}, Hybrid UrQMD \cite{Steinheimer:2012tb}.}
\vspace{0.25cm}
\label{fig:s3_energy} 
\end{minipage}
\end{figure}

Figure \ref{fig:yields_energy} shows the first energy dependence of $\hlt$ production yields in high $\mu_B$ region. An enhanced production of hypernuclei is observed at RHIC BES II energies with respect to LHC energies, which results from the increased baryon density at low energies. The thermal model \cite{Andronic:2010qu} predicts the trend while can not quantitatively describe the yields. Again, we investigate the strangeness population factor versus collision energy, which removes the absolute difference between $\Lambda$ and $p$ yields versus beam energy. Figure \ref{fig:s3_energy} (right) shows the energy dependence of $S_3$ from RHIC to LHC energies. A hint of increasing $S_3$ from $\snn$= 3 GeV to 2.76 TeV is observed. None of the shown models \cite{Zhang:2009ba,Steinheimer:2012tb,Andronic:2010qu} in Fig. \ref{fig:s3_energy} (right) can describe the $S_3$ data quantitatively. We are looking forward to further developments from theory communities.

\subsection{Measurements of hypernuclei lifetimes and branching ratio}
\label{sec1}
\begin{figure}[h]
\centering
\includegraphics[height=3.94cm]{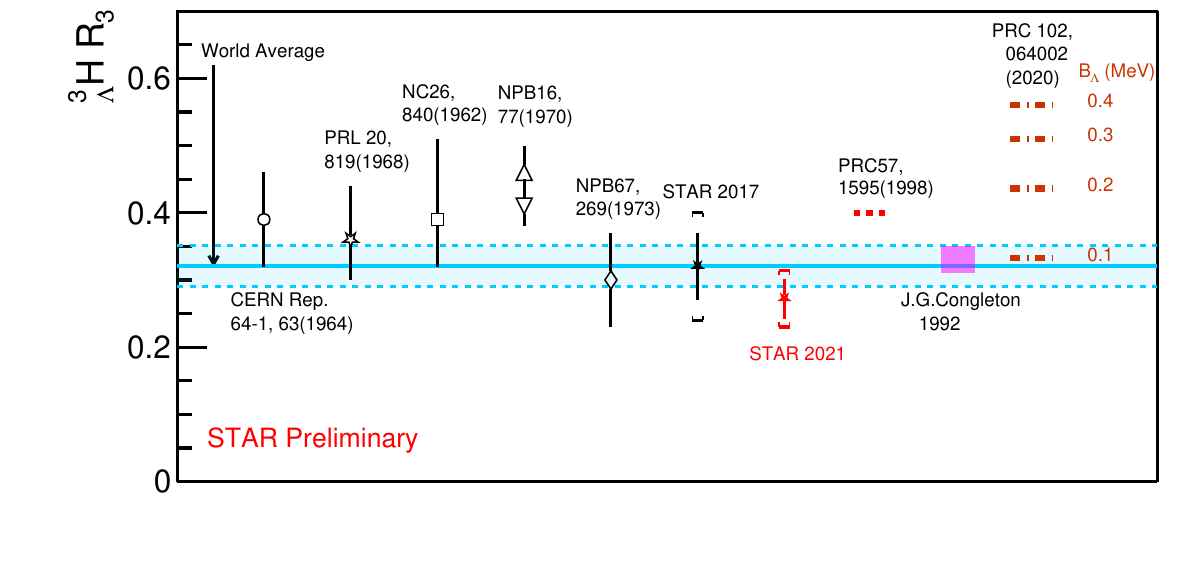}
\includegraphics[height=3.94cm]{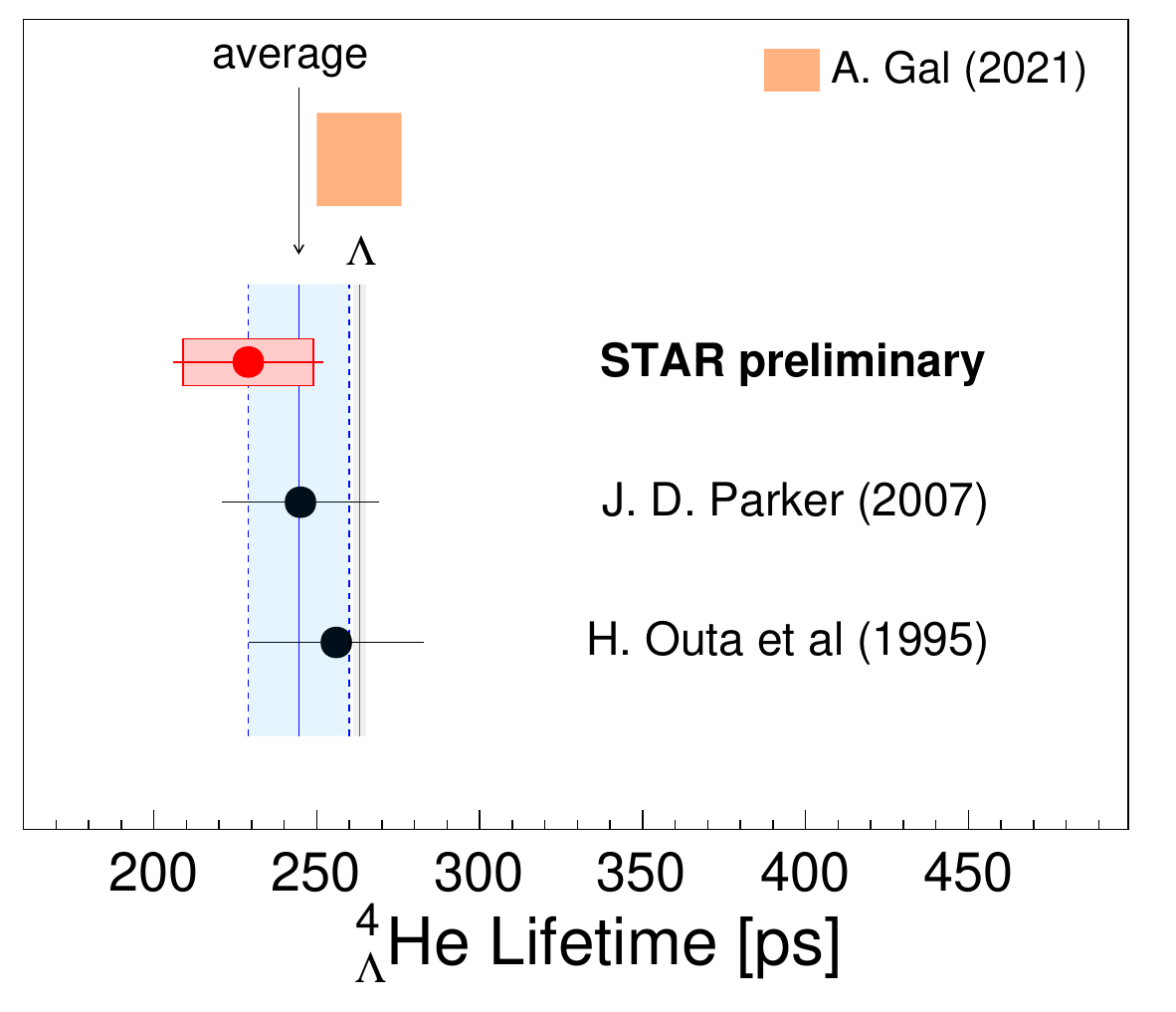}
\caption{The left plot shows the $R_3$ of $^{3}_{\Lambda}{\rm H}$, and the right plot shows the lifetime of $^{4}_{\Lambda}{\rm He}$. The red solid star (left) and circle (right) markers are the STAR new measurements. The updated world average values are indicated as blue bands.}
\label{fig:prop} 
\end{figure}

The $\hlt$ branching ratio $R_3$ is defined as $R_3=\frac{B.R.( \hlt\rightarrow {\rm ^{3}He}\pi^{-})}{B.R.(\hlt\rightarrow dp\pi)+B.R.( \hlt\rightarrow {\rm ^{3}He}\pi)}$.
A recent model calculation \cite{Hildenbrand:2020kzu} predicts that $\hlt$ $R_3$ is sensitive to its binding energy, $B_{\Lambda}$, which is directly connected to $Y$-$N$ interaction strength. 
Hypernuclei lifetime can be extracted by measuring hypernuclei yields as a function of $L/\beta\gamma$ where L is its decay length \cite{STAR:2021orx}. The new measurement on $\hlt$ $R_3$ from STAR is highlighted as the red solid star in Fig. \ref{fig:prop} (left) with the updated world average value $R_3=0.32\pm0.03$ shown as the blue band.
Figure \ref{fig:prop} (right) is the lifetime of $\hel$, where the STAR new result is indicated by the red circle. The updated world average value is consistent with the calculation based on isospin rule \cite{Gal:2021ulo}.

\section{Summary}
\label{sec:sum}
In summary, we present a series of measurements on hypernuclei production and intrinsic properties utilizing high statistics data collected during the BES II program at STAR. We report new measurements on $^{4}_{\Lambda}{\rm He}$ lifetime and $^{3}_{\Lambda}{\rm H}$ $R_3$. The kinematic and centrality dependence of $^{3}_{\Lambda}{\rm H}$ production yields and $S_A$ in 3 GeV Au+Au collisions are presented. Energy dependence of $^{3}_{\Lambda}{\rm H}$ yields and $S_A$ in the mid-rapdity from 3-27 GeV are also reported. Our measurements support the coalescence mechanism of hypernuclei production in heavy-ion collisions. Within the uncertainties, no obvious kinematic or centrality dependence of $S_3$ is observed in 3 GeV Au+Au collisions. Hopefully, our measurements will set strong constraints on hypernuclei internal structures and inspire new insights into the strength of $Y$-$N$ interaction. 

%
\bibliography{ref.bib}
\end{document}